\documentclass[sigconf]{acmart}

\AtBeginDocument{%
  \providecommand\BibTeX{{%
    \normalfont B\kern-0.5em{\scshape i\kern-0.25em b}\kern-0.8em\TeX}}}

\setcopyright{acmcopyright}
\copyrightyear{2022}
\acmYear{2022}
\acmDOI{3503161.3547762}

\usepackage{multirow}

\copyrightyear{2022}
\acmYear{2022}
\setcopyright{acmcopyright}\acmConference[MM '22]{Proceedings of the 30th ACM
International Conference on Multimedia}{October 10--14, 2022}{Lisboa, Portugal}
\acmBooktitle{Proceedings of the 30th ACM International Conference on Multimedia
(MM '22), October 10--14, 2022, Lisboa, Portugal}
\acmPrice{15.00}
\acmDOI{10.1145/3503161.3547762}
\acmISBN{978-1-4503-9203-7/22/10}

\begin{document}

\title{FedMed-ATL: Misaligned Unpaired Cross-Modality Neuroimage Synthesis via Affine Transform Loss}
\author{Jinbao Wang}
\authornote{Equally contribute to this work}
 \affiliation{%
  \institution{Southern University of Science and Technology}
  \country{Shenzhen, China}}
 \email{linkingring@163.com}
 
 \author{Guoyang Xie}
 \authornotemark[1]
 \affiliation{%
  \institution{Southern University of Science and Technology}
  \country{Shenzhen, China}}
       \affiliation{%
  \institution{University of Surrey}
  \country{Guildford GU2 7XH, UK}
  }
 \email{xiegy2019@mail.sustech.edu.cn}

 \author{Yawen Huang}
 \authornotemark[1]
 \affiliation{%
  \institution{Tencent Jarvis Lab}
  \country{Shenzhen, China}}
 \email{yawenhuang@tencent.com}
 
  \author{Yefeng Zheng}
 \affiliation{%
  \institution{Tencent Jarvis Lab}
  \country{Shenzhen, China}}
 \email{yefengzheng@tencent.com}
 
   \author{Yaochu Jin}
 \affiliation{%
  \institution{Bielefeld University}
  \country{33619 Bielefeld, Germany}
  }
   \affiliation{%
  \institution{University of Surrey}
  \country{Guildford GU2 7XH, UK}
  }
 \email{yaochu.jin@surrey.ac.uk}
 
 \author{Feng Zheng}
 \authornotemark[2]
 \affiliation{%
  \institution{CSE and RITAS, Southern University of Science and Technology}
  \country{Shenzhen, China}}
 \email{f.zheng@ieee.org}

\ccsdesc[500]{Computing methodologies~Computer vision tasks}



\begin{abstract}
  The existence of completely aligned and paired multi-modal neuroimaging data has proved its effectiveness in the diagnosis of brain diseases. However, collecting the full set of well-aligned and paired data is impractical, since the practical difficulties may include high cost, long time acquisition, image corruption, and privacy issues. Previously, the misaligned unpaired neuroimaging data (termed as MUD) are generally treated as noisy label. However, such a noisy label-based method fail to accomplish well when misaligned data occurs distortions severely. For example, the angle of rotation is different. In this paper, we propose a novel federated self-supervised learning (FedMed) for brain image synthesis. An affine transform loss (ATL) was formulated to make use of severely distorted images without violating privacy legislation for the hospital. We then introduce a new data augmentation procedure for self-supervised training and fed it into three auxiliary heads, namely auxiliary rotation, auxiliary translation and auxiliary scaling heads. The proposed method demonstrates the advanced performance in both the quality of our synthesized results under a severely misaligned and unpaired data setting, and better stability than other GAN-based algorithms. The proposed method also reduces the demand for deformable registration while encouraging to leverage the misaligned and unpaired data. Experimental results verify the outstanding performance of our learning paradigm compared to other state-of-the-art approaches. 
 
\end{abstract}

\keywords{misaligned unpaired neuroimaging data, unsupervised learning, cross-modality neuroimage synthesis, federated learning}

\maketitle

\section{Introduction}
\begin{figure}[th]
	\centering
    \includegraphics[width=0.85\linewidth]{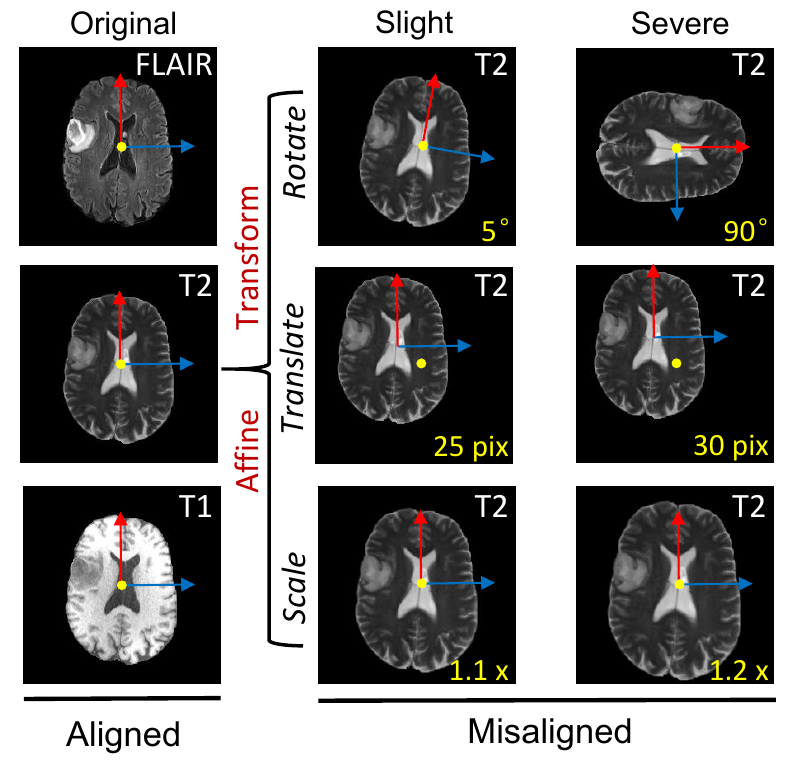}
	\caption{Illustration of multi-modal images, including slightly and severely misaligned cases, paired and unpaired cases. Slight noise denotes the angle ranges in [-3$^{\circ}$, 3$^{\circ}$], the translation ranges in [-15, +15] pixels and the scaling ratio ranges in [0.9, 1.1]. Severe noise denotes the angle ranges in [-90$^{\circ}$, 90$^{\circ}$], the translation ranges in [-30, +30] pixels and the scaling ratio ranges in [0.9, 1.2], respectively
	}\label{fig:severe-misaligned}
\end{figure}

The majority of existing medical datasets \cite{Aljabar2011ACM, Siegel2019CancerS2, Bakas2017BrainlesionGM}, especially for neuroimaging data, are high-dimensional and heterogeneous. For instance, positron emission tomography (PET) and magnetic resonance imaging (MRI) are the imaging techniques to measure the information of organs for auxiliary diagnosis or monitor treatment. The paired/registered multi-modal data provide the complementary information to investigate certain pathologies or neurodegenerations. However, it is often not feasible to acquire complete paired and aligned multi-modal neuroimaging data due to: 1) collecting multi-modal neuroimaging data is very costly; 2) many medical institutions cannot share their data, considering that medical data are especially restricted to the local regulations, despite the identifiable information can be removed for protecting the privacy of patients. 3) patients' motions may result in severe misaligned neuroimaging data. 4) state-of-the-art deformable registration algorithms require tens of minutes to hours for a pair of scans. Hence, most of neuroimaging data are dispersed into different hospitals with misaligned and unpaired property. To solve the above problem, we investigate such data in a realistic scenario. Specifically, we divide the misaligned phenomenon into three parts, severe rotation, severe translation and severe rescaling, which can be referred from the second row of Figure~\ref{fig:severe-misaligned}. The misaligned data setting is similar to RegGAN~\cite{kong2021breaking}, while we set more severe misalignment than that in RegGAN. In particular, the angle rotation setting, the translation setting and the rescaling setting in RegGAN are restricted in [-5, +5] degrees, [-25, +25] pixels and [0.9, 1.1] ratios, respectively. We set the angle rotation, the translation and the rescaling range in [-90, +90] degrees, [-30, +30] pixels, and [0.9, 1.2] ratios, respectively. The detialed simulation about the distribution settings of misaligned unpaired imaging data (MUD) can be referred to Section~\ref{sec:experiments}.

Deformable registration is a fundamental task for solving misaligned medical data. The purpose of deformable registration is to establish a dense and non-linear correspondence between a pair of images, such as 3D misaligned multi-modal neuroimaging data. Traditional methods \cite{avants2011reproducible, klein2009evaluation} align each voxel with a similar appearance by enforcing constraints on the registration mapping. Most of the traditional deformable registration methods can obtain high accuracy. However, they require very huge computation resources resulting in running slowly in practice. Guha et al. \cite{balakrishnan2019voxelmorph} adopt a CNN-based method to speed up the registration procedure. The work in \cite{balakrishnan2019voxelmorph} parameterizes the original deformable registration method as a CNN and optimizes the parameters of CNN by a set of paired but misaligned images. However, the doctors need to verify the effect of deformation algorithms by themselves. This approach inevitably results in a huge amount of labor work and long checking time. Therefore, the emerging issue is to effectively explore MUD rather than using deformable registration to facilitate brain image synthesis. Similar to FedMed-ATL, RegGAN~\cite{kong2021breaking}
forms the misaligned images as the noisy label and utilizes the correction loss to minimize the error resulting from misalignment. However, the assumption of RegGAN \cite{kong2021breaking} is that the neuroimaging data have slight distortions. In this case, RegGAN can convert the slight-distorted neuroimaging data as the noisy label. But in reality, some of the misaligned neuroimaging data inevitably meet with severe distortions. The proposed FedMed-ATL eliminates the requirements of RegGAN and treats MUD as the multi-view data augmentation for the discriminator training, which can significantly mitigate the mode collapse due to the side effect from MUD. Inspired by \cite{chen2020reusing}, FedMed-ATL reuses the discriminator as the encoder. More augmented views of MUD are generated by the Affine Transform Module, and then fed into the designed auxiliary heads, including auxiliary rotation head, auxiliary translation head and auxiliary rotation head. This action aims to strengthen the discriminator to distinguish the real and fake samples. We heuristically prove that FedMed-ATL outperforms RegGAN significantly in a severe distortion setting. Figure~\ref{fig:pipeline} provides detailed information.

Another issue for multi-modality brain image synthesis is data isolation and privacy concern. Although the collaborative research can be conducted by multiple hospitals, they may undertake integrated analysis rather than sharing research data with each other. Lately, a large amount of effort has been made to facilitate the availability of medical data without violating the privacy issue. Federated Learning (FL) is one of the popular approaches. FL is a decentralized approach where local clients train their local models without transmitting data to a central server, and the global model aggregates the gradients from clients \cite{mcmahan2017communication}. In addition, FL with GANs has witnessed some pilot progress on image synthesis \cite{chen2020gs,augenstein2019generative, DBLP:journals/corr/abs-2106-09246}. For example, DP-FedAvg-GAN \cite{augenstein2019generative} trains GANs with the differential privacy-preserving algorithm, which clips the gradients to bound sensitivity and adds calibrated random noise to introduce stochasticity. FedMed-GAN~\cite{xie2022fedmedgan} is similar to our work, however, they assume that all of the neuroimaging data are very clean, i,e, well-aligned in their experimental settings. FedMed-ATL assumes that all neuroimaging data from each hospital are MUD. We found that FedMed-ATL surpasses FedMed-GAN when each client's data is MUD. In addition, we also make a more comprehensive consideration in a realistic scenario and gives more assumptions for all possible data distribution settings and data misaligned conditions. For example, we assume that some hospitals' data are misaligned but paired, and the others are well aligned and paired. In other words, we need to consider both supervised and unsupervised GAN integrated into federated learning. In this case, we can find that the performance of FedMed-ATL is better than FedMed-GAN.  

Our contributions can be summarized as follows:
\begin{itemize}
    \item To the best of our knowledge, FedMed-ATL is the first algorithm to make use of the severely misaligned unpaired data (MUD) to facilitate the multi-modality brain image synthesis. FedMed-ATL reduces the demand for the deformable registration when neuroimaging data meet severe distortions.
   \item Inspired by self-supervised learning, FedMed-ATL treats MUD as a multi-view data augmentation procedure rather than noisy data. We propose three auxiliary losses to the discriminator, including auxiliary rotation, auxiliary translation and auxiliary rescaling losses (termed as the Affine Transform Loss). The proposed method shows good quality of the synthesized results under a severely misaligned and unpaired data setting, and better stability than other Medical-GAN-based algorithms. 
   \item FedMed-ATL encourages to realize the usage of those misaligned and unpaired data when compared with RegGAN, like 90$^{\circ}$ rotation and 20$\%$ translation. In addition, FedMed-ATL exceeds FedMed-RegGAN by a big margin when severe distortions occur in both misaligned and unpaired neuroimaging data.
\end{itemize} 

\section{Related Work}
\subsection{Cross-Modality Medical Image Synthesis}
The existing medical image-to-image translation~\cite{jiang2019synthesize,ren2021segmentation,kong2021breaking} has demonstrated their considerable prospects for both research and clinical analysis. Of these methods, supervised GANs are still the mainstream for cross-modality neuroimaging data synthesis~\cite{Wang2018LocalityAM,Dar2019ImageSI,Sharma2020MissingMP,Yu2020SampleAdaptiveGL,Yu20183DCB,Zuo2021DMCFusionDM}. 
However, synthesizing in a supervised manner requires paired data for training, which is difficult to implement in practice. 
To solve this problem, both semi-supervised and unsupervised methods are then launched to eliminate the need of paired data. 
Guo~\textit{et al.}~\cite{Guo2021AnatomicAM} leverage a lesion segmentation network as a teacher to guide the generator by using unpaired training data. Shen~\textit{et al.}~\cite{Shen2021MultiDomainIC} and Zhou~\textit{et al.}~\cite{Zhou2021AnatomyConstrainedCL} also utilize the high-level tasks to guide the cross-modality image synthesis. Huang~\textit{et al.}~\cite{Huang2020SuperResolutionAI,Huang2020MCMTGANMC} make fully use of unpaired cross-modality data and project them into a common space. The attributed features from the common space bring great helpful to synthesize the missing target modality data. Kong \textit{et al.} \cite{kong2021breaking} introduce a new I2IT model called RegGAN, which converts the unsupervised I2IT task into a supervised I2IT with noisy labels. However, RegGAN cannot deal with the severe distortion, which probably happens in the realistic scenarios.
\subsection{Medical Image Deformable Registration}
There is a lot of work for medical image registration, including statistical parametric mapping~\cite{ashburner2000voxel}, elastic deformation model~\cite{bajcsy1989multiresolution, shen2002hammer}, B-spline~\cite{rueckert1999nonrigid}, Demons~\cite{thirion1998image, pennec1999understanding} and discrete methods~\cite{dalca2016patch}. The work in~\cite{arar2020unsupervised} applies an image-to-image translation network to preserve the geometric property, which may be lost from deformable regularization. Recently, deep learning based methods are adapted to learn the parameters of deformation field. In total, deep learning based methods can be classified as supervised learning methods~\cite{Roh2017SVFNetLD,Fan2018AdversarialSN} and unsupervised learning methods~\cite{Wang2020DeepFLASHAE,Liu2020BilevelPF}. The supervised learning methods require the ground truth deformation fields as the training dataset but the unsupervised methods do not need the ground truth. In addition, medical image registration methods can also be categorized as the displacement field~\cite{Heinrich2020HighlyAA,Yang2022LDVoxelMorphAP} and the diffeomorphic methods~\cite{Mok2020FastSD,Dalca2019UnsupervisedLO,Jia2022LearningAM,Thorley2021NesterovAA}. The displacement methods can directly obtain the deformation field via CNN model. While the diffemorphic methods desire to guarantee the diffemorphic property by computing the diffemophic deformation field. Both of them aims to obtain better registration results instead of image synthesis. FedMed-ATL aims to utilize both misaligned and unpaired brain image data for synthesis without a registration procedure. 

\subsection{Federated Learning}
Data isolation and privacy concerns are the fundamental problems which blocks a large-scale and multi-institute neuroimaging research. Federated learning, as a privacy-preserving decentralization strategy, allows clients train their own models without data communication to a central server by aggregating client progresses to update a global model~\cite{DBLP:conf/aistats/McMahanMRHA17,DBLP:conf/icml/YurochkinAGGHK19,DBLP:conf/mlsys/LiSZSTS20,DBLP:conf/iclr/WangYSPK20}. FedAvg \cite{DBLP:conf/aistats/McMahanMRHA17} combines local stochastic gradient descent (SGD) on each client with a server that performs model averaging. Yurochkin \textit{et al.} \cite{DBLP:conf/icml/YurochkinAGGHK19} developed a Bayesian non-parametric framework for federated learning with neural networks. 
FedProx \cite{DBLP:conf/mlsys/LiSZSTS20} provides a generalized and re-parametrized FedAvg that addresses the challenges of heterogeneity both theoretically and empirically.
FedMA \cite{DBLP:conf/iclr/WangYSPK20} constructs the shared global model in a layer-wise manner by matching and averaging hidden elements with similar feature extraction signatures. Ho. wever, directly incorporating the generative adversarial framework into the federated learning is challenging, due to the cost functions may not converge using federated gradient aggregation in a min-max setting between the discriminator and the generator~\cite{DBLP:conf/iclr/AugensteinMRRKC20,DBLP:conf/nips/ChenOF20,DBLP:journals/corr/abs-2106-09246}. 
DP-FedAvg-GAN \cite{DBLP:conf/iclr/AugensteinMRRKC20} provides a differential privacy-preserving algorithm, which clips the gradients to bound sensitivity and adds calibrated random noise to introduce stochasticity. GS-WGAN \cite{DBLP:conf/nips/ChenOF20} enables the release of a sanitized version of sensitive data while maintaining stringent privacy protections. 
GS-WGAN is capable to distort gradient information, which allows to train a deeper model with more informative samples. 
Despite these methods demonstrate their resonable results on different tasks, the performance on cross-modality image synthesis is untapped, let alone the theory or empirical explanations for the convergence.
In addition, the proportion of paired and unpaired data for each client (hospital) and the long-tail data distribution problems are all neglected.

\begin{figure*}[t]
	\centering
    \includegraphics[width=0.85\linewidth]{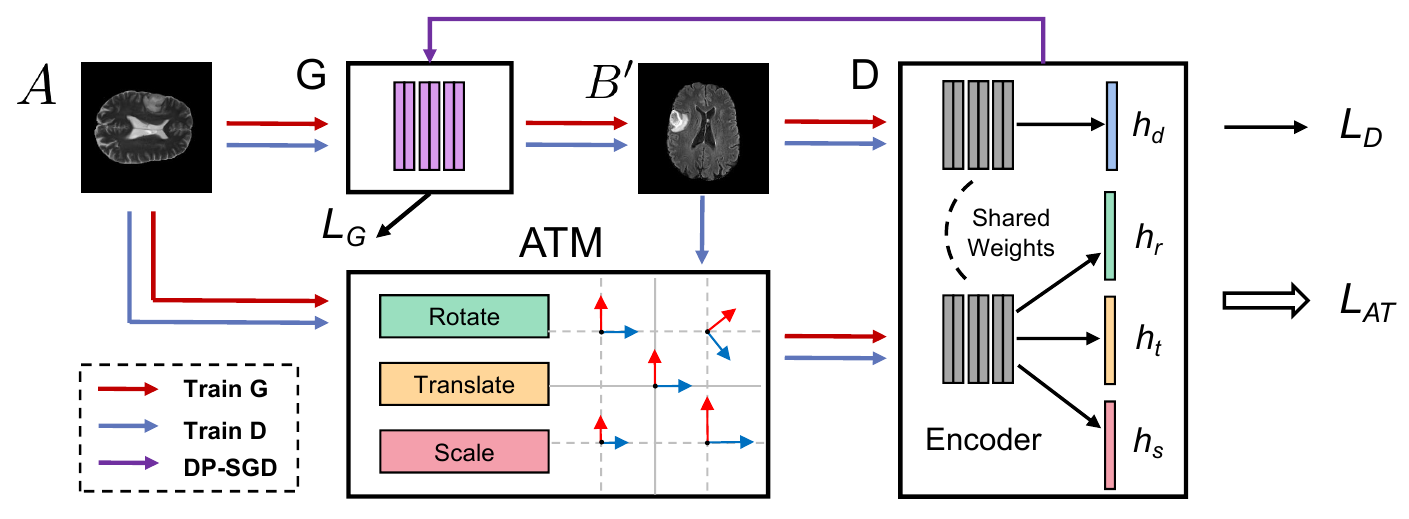}
	\caption{
		The pipeline of FedMed-ATL. The generator G generates B-modal images $B'$ from A-modal. The discriminator D distinguishes $B'$ from real or fake. The affine transform module (ATM) achieves image transform. The functions of $L_{G}$, $L_{D}$, and $L_{AT}$ are the generator loss, the discriminator loss, and the affine transform loss, respectively. To train G and D, the red line and blue line describe the data flow. Note that input A-model images are fed into G and ATM for training G, but both input A-model images and generated B-model images $B'$ are fed into D. In ATM, three transform blocks (Rotate, Translate, Scale) provide a rich data resource. For federated learning, we add DP-SGD~\cite{abadi2016deep} (purple line) to avoid privacy leakage. 
	}\label{fig:pipeline}
\end{figure*}

\begin{figure}[thbp]
    \centering
    \includegraphics[width=0.7\linewidth]{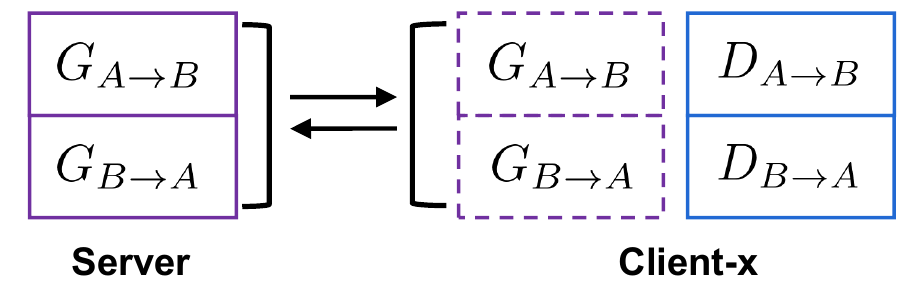}
	\caption{
		The federated setting of FedMed-ATL model. $G$ denotes the shared generator and $D$ denotes the discriminator. The discriminator of each client does not participate into the aggregation process. Only the generator of each client join into the aggregation process. 
	}\label{fig:fed_model}
\end{figure}

\section{FedMed-ATL}\label{sec:fedmed_atl}
\subsection{Federated Model Setup}\label{fed_model_setup}
The federated setting of our generator and discriminator is described in Figure~\ref{fig:fed_model}. We employ CycleGAN \cite{zhu2017unpaired} as our baseline model.
CycleGAN \cite{zhu2017unpaired} owns two generators $G_{A \leftrightarrow B}$, discriminators $D_{A \leftrightarrow B}$. Specifically, $G_{A \rightarrow B}$ generates B-modal images from A-modal samples. $D_{A \rightarrow B}$ distinguishes whether the generated B-modal data from A-modal sample is fake. The federated setting of CycleGAN is to locate two generators ($G_{A \rightarrow B}$, $G_{B \rightarrow A}$) into the servers. 
On the other hand, the generators are separately aggregated into the server's generators. Then, the server sends its generators into different hospitals. The discriminators ($D_{A \rightarrow B}$, $D_{B \rightarrow A}$) of each client are independent.

\subsection{Misaligned Unpaired Data Distribution}\label{sec:data_settings}
\begin{figure}[thbp]
    \centering
    \includegraphics[width=0.85\linewidth]{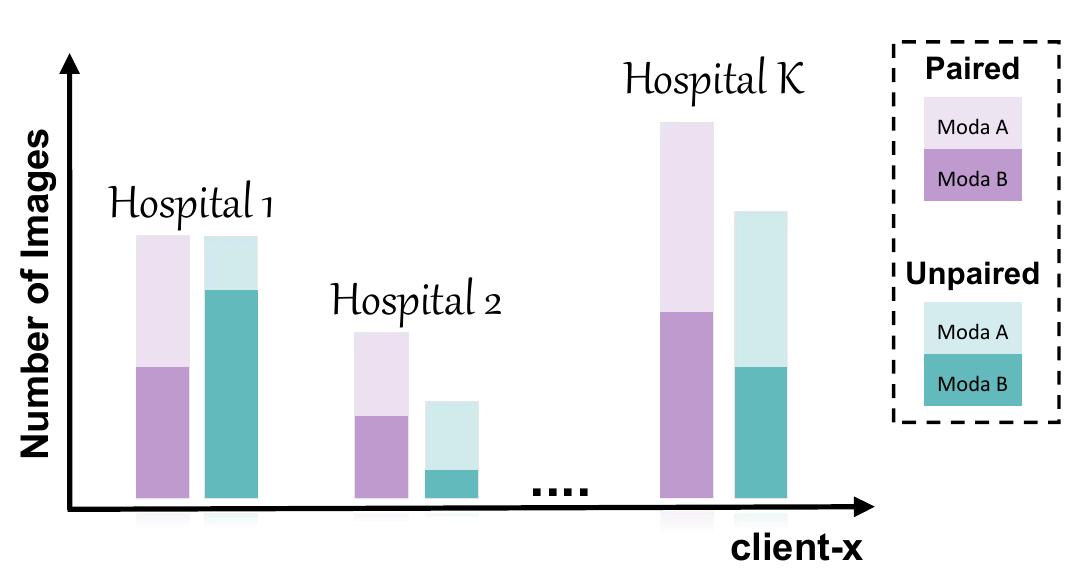}
	\caption{
		 Distribution of misaligned unpaired data (MUD). For instance, Hospital 1 has paired data with misaligned attributes, while Hospitals of 2-K have MUD in the most real-life case. 
	}\label{fig:data-settings}
\end{figure}

We address the task of federated self-supervised learning for MUD by formulating several realistic settings. To simulate the real data distribution as much as possible, we adopt the following settings as shown in Figure \ref{fig:data-settings}.

Firstly, we divide ready-made data into several clients, where each client has its own private and unique data. Meanwhile, these clients (hospitals) contain unbalanced multi-modal images, i.e. having different numbers of patients in each client or having different numbers of images in each modality. Secondly, without losing generality, we explore both paired and unpaired cases in our experiments. For example, we randomly select the multi-modal slices from the same patient as paired data like in Hospital 1 or select them from patients with different IDs as unpaired data like in Hospital 2. After that, we randomly transform the input images at a certain threshold of rotation, translation and scaling, and there is the majority of misaligned images in the constructed training set.  
\subsection{Network Architectures}\label{sec:architecture}
\textbf{Generator}
In general, the architecture of generator could be encoder-decoder style. The performance of FedMed-ATL is not being affected by the specific generator's architecture. In Figure~\ref{fig:pipeline}, FedMed-ATL use U-Net~\cite{Ronneberger2015UNetCN} as the generator.

\textbf{Discriminator}
Inspired by Chen \textit{et.al.}~\cite{chen2020reusing}, FedMed-ATL reuses the original discriminator as the encoder. The architecture of discriminator is similar to CycleGAN~\cite{zhu2017unpaired}, MUNIT~\cite{DBLP:conf/eccv/HuangLBK18} and UNIT~\cite{DBLP:conf/nips/LiuBK17} except the last layer. FedMed-ATL replaces the last layer of the discriminator with a two-layer MLP (termed as discriminator head). In Figure~\ref{fig:pipeline}, $h_{d}$ represents the discriminator head. The input and output dimensions of two layers are (512, 128) and (128, 1), respectively. The role of the discriminator head is to distinguish the real or fake samples.


\textbf{Affine Transform Module (ATM)}\label{sec:atm}
ATM contains three operators, namely rotation, translation and rescaling. As for rotations, the rotation degree is chosen in [0$^{\circ}$, 90$^{\circ}$, 180$^{\circ}$, 270$^{\circ}$]. As for translation, the x-axis and y-axis translation is chosen in [(-30, -30), (-30, 30), (30, -30), (30, 30)]. As for rescaling, the scaling ratio is chosen in [0.9, 1.1, 1.2]. We conduct a comprehensive ablation study on the number of views for each operation. The number of views is a range in [1, 2, 4]. For instance, if the number of views is 4, FedMed-ATL picks 4 augmented image data from rotation, translation and rescaling, and then feeds them into auxiliary rotation head, auxiliary translation head and auxiliary rescaling head, respectively. In addition, when FedMed-ATL trains the generators, only the real samples are fed into ATM. The real and fake samples are fed into the ATM during the discriminator training procedure. This treatment encourages the generator to induce more 'really misaligned' samples for the discriminator to avoid overfitting. 

\textbf{Auxiliary Rotation Head}
The role of auxiliary rotation head ($h_{r}$ in Figure~\ref{fig:pipeline}) is to distinguish the rotation angle of the neuroimaging data, i.e, 0$^{\circ}$, 90$^{\circ}$, 180$^{\circ}$ and 270$^{\circ}$. Before going into the auxiliary rotation head, FedMed-ATL needs to feed them into the encoder. This encoder is shared with the other heads, including the discriminator head, translation head, and scaling head. In Equation~\ref{eq:rotation_loss}, $x$ are the real samples when FedMed-ATL update the parameters of generators. In Equation~\ref{eq:rotation_loss}, $x$ denotes both the real sample and fake samples (the generated samples) when FedMed-ATL updates the parameters of discriminators. $P_{h_{r}}(R | x^{r})$ is the predictive distribution of auxiliary rotation head over the angles of rotation of the sample. The intuition of auxiliary rotation loss is to enable the discriminator to access more severe rotated samples. Since the rotation head and discriminator head share the same encoder, i.e., these two heads own similar features. In this case, even if the discriminator meets with the severely rotated samples, the auxiliary rotation loss can help the discriminator to distinguish real or fake samples without being affected by severe rotation.

\textbf{Auxiliary Translation Head}
The role of auxiliary translation head ($h_{t}$ in Figure~\ref{fig:pipeline}) is to distinguish the translation distance of the neuroimaging data, i.e,  [(-30, -30), (-30, 30), (30, -30), (30, 30)]. Similarly, $x$ denotes the real sample when FedMed-ATL updates the parameters of generators. 
$P_{h_{t}}(T | x^{t})$ is the predictive distribution of auxiliary translation head over the translation direction and distance of the sample. The aim of auxiliary translation loss is similar to the auxiliary rotation loss. In this case, the discriminator is able to access more severe translated samples. The auxiliary translation loss can support the discriminator to distinguish real or fake samples without being affected by severe translation.

\textbf{Auxiliary Rescaling Head}
The role of auxiliary rescaling head ($h_{s}$ in Figure~\ref{fig:pipeline}) is to distinguish the rescaling ratio of neuroimaging data, i.e,  [0.9, 1.1, 1.2]. 
Similar to the above two heads, $P_{h_{s}}(S | x^{s})$ denotes the predictive distribution of auxiliary rescaling head over the scaling ratio of the sample. 
The auxiliary scaling loss can further help the discriminator to distinguish the synthesized samples without being affected by severe rescaling.

\subsection{Loss Function}\label{sec:loss}
\textbf{AT Loss} Three affine transform modes, namely rotation, translation, and scaling, are defined as follows.
\begin{equation}\label{eq:rotation_loss}
    L_{rot} = \lambda_{rot}\cdot \mathbb{E}[\log\, P_{h_{r}} (R=r | x^{r})]
\end{equation}
\begin{equation}\label{eq:translation_loss}
    L_{trans} = \lambda_{trans}\cdot \mathbb{E}[\log\,P_{h_{t}} (T=t | x^{t})]
\end{equation}
\begin{equation}\label{eq:scaling_loss}
    L_{scale} = \lambda_{scale}\cdot \mathbb{E}[\log\,P_{h_{s}} (S=s | x^{s})]
\end{equation}

\textbf{GAN and Cycle Loss}
Equation~\eqref{eq:gan_loss} is the loss function of GAN~\cite{DBLP:conf/cvpr/IsolaZZE17}, where $G$ is the generator that generates B-modal images from A-modal images, and $D$ is the discriminator that distinguishes whether B-modal images generated from A-modal samples is fake. 

\begin{equation} \small
\begin{aligned}
     \min_G \max_G L_{adv}(G,D) =& \mathbb{E}_y[\log(D(y))]\\
     & + \mathbb{E}_x [\log(1-D(G(x)))]
\end{aligned}\label{eq:gan_loss}
\end{equation}
Equation~\eqref{eq:cycle_loss} is Cycle loss function~\cite{zhu2017unpaired}, where $F$ is the generator that generate A-modal image from B-modal images.

\begin{equation} \small
\begin{aligned}
     \min_G \max_G L_{cyc}(G,F)= & \mathbb{E}_x [||F(G(x)) -x ||_1]  \\
     & +  \mathbb{E}_y [||G(F(y))-y||_1]
\end{aligned}\label{eq:cycle_loss}
\end{equation}

\textbf{Total Loss}
Overall, the total loss function is defined as:
\begin{equation} 
\begin{aligned}
     L_{total} = & \lambda_{rot} \cdot L_{rot} + \lambda_{trans} \cdot L_{trans} \\
     & + \lambda_{scale} \cdot L_{scale} + \lambda_{adv} \cdot L_{adv} \\
     & + \lambda_{cyc} \cdot L_{cyc}.   
\end{aligned}\label{eq:total_loss_1}
\end{equation}

The weight $\lambda$ of various loss functions will be introduced in the implementation part of Section~\ref{sec:experiments}.

\begin{table*}[t]
	\centering \small
	\setlength{\tabcolsep}{4mm}
	\caption{Comparison results of modal generation (from PD to T2) on the IXI dataset.}
    \begin{tabular}{l|c|c|c|c|c|c}
    \hline
    \multirow{2}{*}{\textbf{Method}}   & \multicolumn{3}{c|}{\textit{Slight Noise}}  & \multicolumn{3}{c}{\textit{Severe Noise}} \\\cline{2-7}
      & \textbf{MAE} $\downarrow$    & \textbf{PSNR} $\uparrow$    & \textbf{SSIM} $\uparrow$      & \textbf{MAE} $\downarrow$    & \textbf{PSNR} $\uparrow$    & \textbf{SSIM} $\uparrow$    \\\hline
    FedMed-C (+reggan)   &  0.0294      &  \textbf{ 26.0068}      &   \textbf{0.9614 }       &   0.0302  &  24.5915       &  0.9477           \\\hline
    FedMed-C  &  0.0385      &  24.9238  &   0.9545  &  0.0295   &  \textbf{25.0715}  &  \textbf{0.9555}    \\\hline
    FedMed-C-ATL  &   \textbf{0.0279}   &  25.0844    & 0.9560    & \textbf{0.0290} &  24.7132     & 0.9517           \\\hline 
    \end{tabular}
    \label{ex:baseline_ixi}
\end{table*}

\begin{table*}[t]
	\centering \small
	\setlength{\tabcolsep}{4mm}
	\caption{Comparison results of modal generation (from T1 to FLAIR) on the BraTS2021 dataset.}
    \begin{tabular}{l|c|c|c|c|c|c}
    \hline
    \multirow{2}{*}{\textbf{Method}}   & \multicolumn{3}{c|}{\textit{Slight Noise}}  & \multicolumn{3}{c}{\textit{Severe Noise}} \\\cline{2-7}
      & \textbf{MAE} $\downarrow$    & \textbf{PSNR} $\uparrow$    & \textbf{SSIM} $\uparrow$      & \textbf{MAE} $\downarrow$    & \textbf{PSNR} $\uparrow$    & \textbf{SSIM} $\uparrow$    \\\hline
    FedMed-C (+reggan)  &   0.0481     &  19.8538       &   0.8841         &   0.0455     &   20.0179      &  0.8950           \\\hline
    FedMed-C   &  0.0466    &   20.0601   &   0.9003          & 0.0474    & 19.9319     &   0.8928             \\\hline
    FedMed-C-ATL  &   \textbf{0.0447}     &   \textbf{20.3961}      & \textbf{ 0.9044  }           &  \textbf{0.0439}       & \textbf{20.8232  }      &    \textbf{0.9123}        \\\hline
    \end{tabular}
    \label{ex:baseline_brats}
\end{table*}

\section{Experiment}\label{sec:experiments}
\subsection{Federated Settings}\label{sec:experiment_data_settings}
\textbf{IXI}~\cite{Aljabar2011ACM} collects nearly 600 MR images from normal and healthy subjects at three hospitals. The MR image acquisition protocol for each subject includes T1, T2, PD-weighted images (PD), MRA images, and Diffusion-weighted images. In this paper, we only use T1 (581 cases), T2 (578 cases) and PD (578 cases) data to conduct our experiments, and select the paired data with the same ID from the three modes. The image has a non-uniform length on the z-axis with the size of $ 256 \time 256 $ on the x-axis and y-axis. The IXI dataset is not divided into a training set and a test set. Therefore, we randomly split the whole data as the training set (80\%) and the test set (20\%).

\textbf{BraTS2021} \cite{Siegel2019CancerS2, Bakas2017BrainlesionGM} is constructed for analysis and diagnosis of brain disease. The publicly available dataset of multi-institutional and pre-operative MRI sequences is provided, which includes both training data (1251 cases) and validation data (219 cases). Each 3D volume has 155$\times$240$\times$240 size imaged by four sequences: T1, T2, T1ce, and FLAIR. 

\textbf{Data processing} To ensure data validity and diversity, we remove a skull without tissues in a slice, and split the three-dimensional volume and select slices from 50 to 80 on the Z-axis. All of the images in all datasets are cropped into the size of $ 256 \time 256 $ pixels. We define unpaired data as two modal slices (T1, FLAIR) from two volumes (A and B) with the same N-th slice, e.g., unpaired data [T1-A-N, FLAIR-B-N]. By affinely transforming $at(\cdot)$ both two modal images, we obtain a MUD-liked data [$at$(T1-A-N), $at$(FLAIR-B-N)]. In our federated scenario, we first divide the training data (volume) proportionally into 4 clients based on the size of client data [0.4, 0.3, 0.2, 0.1], where each client has its own private and unique ones. After that, we construct MUDs in each client via the generation principle of unpaired data and misaligned data. In the training stage, we select overall 6,000 images from the IXI and BraTS2021 datasets, and each client selects 2,400, 1,800, 1,200, and 600 from the corresponding dataset. For evaluation, we use 3,462 images as our validation data from the IXI dataset, and 6,570 images from the BraTS2021 dataset, respectively. 

\textbf{Metrics} We employ three metrics to evaluate our generator's performance. The first is mean absolute error (MAE):
\begin{equation}\small
    MAE = \frac{1}{nm}\sum_{n}^{i=1}\sum_{m}^{j=1}\left| T_{ij} - G_{ij} \right|,
\end{equation}
where $T_{ij}$ denotes the ground truth neuroimage pixel and $G_{ij}$ denotes the generated neuroimage pixel. The lower value of MAE means the better performance.

The second metric is the peak signal-to-noise ratio (PSNR). PSNR is a function of the mean squared error and better to evaluate the context (edge) detail of neuroimages. The higher PSNR value means the better performance. 
\begin{equation}\small
    PSNR = -10 \: log_{10}\left ( \frac{1}{nm}\sum_{n}^{i=1}\sum_{m}^{j=1} (T_{ij} - G_{ij})^{2} \right )
\end{equation}
The third metric is structural similarity index (SSIM), which is a weighted combination of the luminance, the contrast and the structure. The higher SSIM value means the better performance.
\begin{equation}\small
\mathrm{SSIM}=\frac{\left(2 \mu_{T} \mu_{G}+C_{1}\right)\left(2 \sigma_{T G}+C_{2}\right)}{\left(\mu_{T}^{2}+\mu_{G}^{2}+C_{1}\right)\left(\sigma_{T}^{2}+\sigma_{G}^{2}+C_{2}\right)}
\end{equation}\label{eq:ssim}
The $\mu$ and $\sigma$ in SSIM are the mean value and standard deviation of an image, respectively. $C_{1}$ and $C_{2}$ are two positive constants. We set $C_{1}$ and $C_{2}$ are 0.01 and 0.03, respectively.The values with lower MAE, the higher PSNR and SSIM denote the higher quality of the synthesized images.

\textbf{Implementations} 
In the federated scenario, there are 4 clients in our experiment, where the number of volumes owned by each client is 0.4 0.3, 0.2 and 0.1 respectively. We run 3 rounds for federated training, and each client is trained for 3 epochs. The model aggregation strategy is based on Fed-Avg \cite{mcmahan2017communication}, which aggregates the weight from each client's generator model to the server model according to the data proportion distribution for each client. Specifically, we use the learning rate of $1e-4$, and the batch size of 4. The optimizer is Adam \cite{kingma2014adam}. Its beta1 and beta2 are 0.5 and 0.999, respectively. The weights of GAN loss $\lambda_{adv}$ and Cycle loss $\lambda_{cyc}$ are 1.0, 10.0, respectively. For the affine transform loss (ATL), we set weights of rotation $\lambda_{rot}$, translation $\lambda_{trans}$ and scaling $\lambda_{scale}$ to 1.0 for each model's generator and 0.5 for its discriminator, respectively. In FedMed-ATL differential privacy settings, the level of gradient clip bound, sensitivity, and noise multiplier is fixed to 1.0, 2.0, and 1.07, respectively. In terms of differential privacy setting, the Gaussian noise $\mu$ is set to 1.07, and the standard deviation $\sigma$ is set to 2.0. The clip bound for the back-propagation gradient is 1.0. 

\textbf{Slight and Severe Noise} In our experiments, slight noise denotes the noise level is 3 like RegGAN~\cite{kong2021breaking}. Specifically, slight noise denotes the angle ranges in [-3$^{\circ}$, 3$^{\circ}$], the translation ranges in [-15, +15] pixels and the scaling ratio ranges in [0.9, 1.1]. On the other hand, severe noise denotes the angle ranges in [-90$^{\circ}$, 90$^{\circ}$], the translation ranges in [-30, +30] pixels and the scaling ratio ranges in [0.9, 1.2], respectively.

\begin{table*}[ht]
	\centering \small
	\setlength{\tabcolsep}{4mm}
	\caption{Results of modal generation (from PD to T2) on the IXI dataset.}
    \begin{tabular}{l|c|c|c|c|c|c}
    \hline
    \multirow{2}{*}{\textbf{Method}}   & \multicolumn{3}{c|}{\textit{Slight Noise}}  & \multicolumn{3}{c}{\textit{Severe Noise}} \\\cline{2-7}
      & \textbf{MAE} $\downarrow$    & \textbf{PSNR} $\uparrow$    & \textbf{SSIM} $\uparrow$      & \textbf{MAE} $\downarrow$    & \textbf{PSNR} $\uparrow$    & \textbf{SSIM} $\uparrow$    \\\hline
    FedMed-C        &  0.0385      &  24.9238       &   0.9545    &  0.0295      &  \textbf{25.0715}       &  \textbf{0.9555}            \\\hline
    FedMed-C-AR  &  \textbf{0.0261}      &  \textbf{25.3358}       &   \textbf{0.9577}    & 0.0298     & 24.3385    &  0.9467    \\\hline
    FedMed-C-AT  &  0.0271      &  25.2926       &  0.9569     &  0.0291   &  24.7600    & 0.9509   \\\hline
    FedMed-C-AS  &  0.0297      &  24.7426       &  0.9508     &  0.0316      & 24.0586        &   0.9410       \\\hline
    FedMed-C-ATL-1View  & 0.0311   & 24.7486      &  0.9562      & 0.0331     & 24.2766    &  0.9518      \\\hline
    FedMed-C-ATL-2Views  & 0.0288   & 24.8814      &  0.9562      & 0.0305    &  24.3427   &  0.9488     \\\hline
    FedMed-C-ATL-4Views    &  0.0279   &  25.0844    & 0.9560      & \textbf{0.0290} &  24.7132     & 0.9517        \\\hline
    \end{tabular}
    \label{ex:baseline_ixi_loss}
\end{table*}

\begin{table*}[ht]
	\centering \small
	\setlength{\tabcolsep}{4mm}
	\caption{Results of modal generation (from T1 to FLAIR) on the BraTS2021 dataset.}
    \begin{tabular}{l|c|c|c|c|c|c}
    \hline
    \multirow{2}{*}{\textbf{Method}}   & \multicolumn{3}{c|}{\textit{Slight Noise}}  & \multicolumn{3}{c}{\textit{Severe Noise}} \\\cline{2-7}
      & \textbf{MAE} $\downarrow$    & \textbf{PSNR} $\uparrow$    & \textbf{SSIM} $\uparrow$      & \textbf{MAE} $\downarrow$    & \textbf{PSNR} $\uparrow$    & \textbf{SSIM} $\uparrow$    \\\hline
    FedMed-C        &  0.0466      &   20.0601   &   0.9003       & 0.0474    & 19.9319     &   0.8928             \\\hline
    FedMed-C-AR  &  0.0483      &  19.9217       &  0.8904   & 0.0439     &  20.4555   & 0.9056  \\\hline
    FedMed-C-AT  &  0.0466      &  20.2059       &  0.8944    &  0.0440   &  20.4512   &  0.9042   \\\hline
    FedMed-C-AS  &  0.0459      &  20.0981       &  0.8956    &   \textbf{0.0435}     & 20.6446        &  0.9093        \\\hline
    FedMed-C-ATL-1View  & 0.0451   &   \textbf{20.6151}    &  \textbf{0.9085}     &  0.0439    &  20.8199   &   0.9115   \\\hline
    FedMed-C-ATL-2Views  & 0.0451    &   20.2883    & 0.9012       &  0.0459   &  20.4236   & 0.9047   \\\hline
    FedMed-C-ATL-4Views  & \textbf{0.0447}  &   20.3961      & 0.9044    &  0.0439       & \textbf{20.8232}     &    \textbf{0.9123}     \\\hline
    \end{tabular}
    \label{ex:baseline_brats_loss}
\end{table*}

\subsection{Main Results}

For the notation used in Table~\ref{ex:baseline_ixi}, Table~\ref{ex:baseline_brats}, Table~\ref{ex:baseline_ixi_loss} and Table~\ref{ex:baseline_brats_loss}, M, C and U are the short names for MUNIT~\cite{DBLP:conf/eccv/HuangLBK18}, CycleGAN~\cite{zhu2017unpaired} and UNIT~\cite{DBLP:conf/nips/LiuBK17}, respectively. 

\textbf{FedMed-C} denotes that we only put CycleGAN into our federated settings, which was introduced in Section 3.1 and Section 4.1 by adopting cycle loss~\eqref{eq:cycle_loss} and adversarial loss \eqref{eq:gan_loss}. \textbf{FedMed-C (+reggan)} denotes that FedMed-C adopts the correction loss and registration network described in RegGAN~\cite{kong2021breaking}. \textbf{FedMed-C-ATL} denotes that FedMed-C employs the affine transform loss (ATL), which was introduced in Equation~\eqref{eq:total_loss_1}.

Table~\ref{ex:baseline_ixi} and Table~\ref{ex:baseline_ixi_loss} show the results of the generated T2 from PD in the IXI dataset. Table~\ref{ex:baseline_brats} and Table~\ref{ex:baseline_brats_loss} show the results of the generated FLAIR from T1 in the BraTS2021 dataset.


\subsection{Ablation Study}

\textbf{Analysis of ATL} To evaluate the performance of the affine transform loss, we test rotation, translation and sacling, respectively. 
\textbf{FedMed-C-AR} denotes that FedMed-C just employs the auxiliary rotation loss shown in Equation~\eqref{eq:rotation_loss}. \textbf{FedMed-C-AT} denotes that FedMed-C just employs the auxiliary translation loss given in Equation~\eqref{eq:translation_loss}. \textbf{FedMed-C-AS} denotes that FedMed-C just employs the auxiliary scaling loss provided in Equation~\eqref{eq:scaling_loss}.

\textbf{Analysis of Different Views in ATL} We further investigate the impact of different views on ATL. \textbf{FedMed-C-ATL-1View} denotes that FedMed-C-ATL randomly picks one augmented view from the ATM Module and feeds it into the encoder (discriminator), which is described in Figure~\ref{fig:pipeline} and Section~\ref{sec:atm}. In other words, there is only one augmented view to fed into the auxiliary rotation head, auxiliary translation head, and auxiliary rescaling head, respectively. \textbf{FedMed-C-ATL-2Views} denotes that FedMed-C-ATL randomly picks two augmented views and feeds them into the encoder (discriminator). The specific operation is similar to FedMed-C-ATL-1View, while the number of views is 2. \textbf{FedMed-C-ATL-4Views} denotes that FedMed-C-ATL picks four augmented views and feeds them into the encoder (discriminator). Note that in the previous description, the default way is to pick 4 views if this paper does not mention the number of views explicitly.

\textbf{Key Findings 1}
We compare FedMed-C-ATL with the existing state-of-the-art algorithm, i.e. RegGAN~\cite{kong2021breaking}. We can observe that FedMed-C-ATL outperforms FedMed-RegGAN by a large margin in a severe MUD setting, which achieves 2.4~$\%$ - 4.6~$\%$ absolute improvements in MAE, PSNR and SSIM, respectively. The results demonstrate that FedMed-C-ATL has a better ability to facilitate the quality of the synthesis, even if the input data suffered from a severe distortion. In addition, FedMed-C-ATL achieves absolute improvements of 10~$\%$, 2.7~$\%$ and 2.7~$\%$ in slight MUD setting, when compared with FedMed+RegGAN. The improvement demonstrates the effectiveness of our FedMed-C-ATL for the slight MUD scenario. We also observe that FedMed-C performs slightly better than FedMed-C-ATL in the severe MUD settings on the IXI dataset. The principal reason is that data augmentation of FedMed-C-ATL results in opposite effect, when facing a very clean dataset like IXI. Here, ``clean" does not mean a well-aligned and paired dataset, instead, it means healthy subject. From the results, we found that excessive data augmentations for a clean dataset are meaningless and even harmful for synthesis.

\textbf{Key Findings 2}
To further evaluate the effectiveness of FedMed-C-ATL, we provide an ablation study to verify the usefulness of various auxiliary heads. In Table~\ref{ex:baseline_brats_loss}, it the results show that by adding more heads can improve performance in a severe MUD setting. Specifically, FedMed-C-ATL-4views outperforms FedMed-C-AS by a large margin in a severe MUD setting, which achieves 6.8~$\%$, 1.7~$\%$ and 1.1~$\%$ absolute improvements in MAE, PSNR and SSIM, respectively. From Table~\ref{ex:baseline_ixi_loss}, FedMed-C-ATL-4views surpasses FedMed-C-AS by 1.1 $\%$ and 2.1~$\%$ in MAE, PSNR and SSIM, respectively. 

\textbf{Key Findings 3}
Furthermore, we desire to understand what quantity is suitable for data augmentation. From Table~\ref{ex:baseline_brats_loss} and Table~\ref{ex:baseline_ixi_loss}, we observe that FedMed-C-ATL-4views outperforms FedMed-C-ATL-1view and FedMed-C-ATL-2views in severe MUD settings. For example, FedMed-C-ATL-4views outperforms FedMed-C-ATL-1view by 1.2~$\%$, 2.1~$\%$ and 2.7~$\%$ in MAE, PSNR and SSIM, respectively. It demonstrates that the number of data augmentation plays a positive role in the quality of the synthesized results, especially in severe MUD settings.

\subsection{Visualization}

\begin{figure}[thbp]
    \centering
    \includegraphics[width=0.8\linewidth]{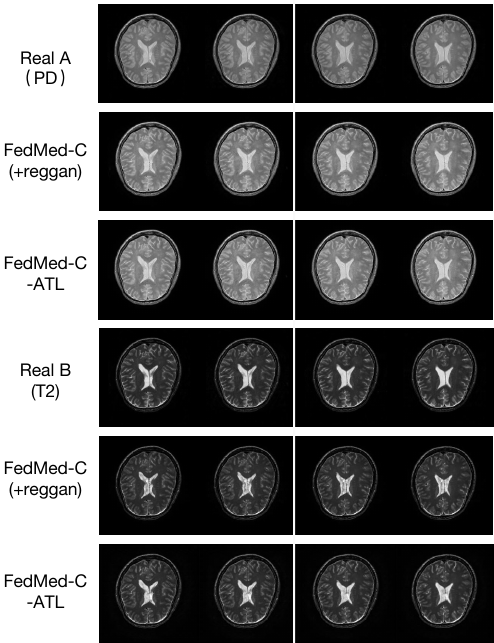}
	\caption{Visualization of brain images generated by FedMed-C~(+reggan) and FedMed-C-ATL on the IXI validation set. The model is trained on the severe noise scheme. The row images represent two volume examples of two near slices. 
	}\label{fig:vis-ixi}
\end{figure}

\begin{figure}[thbp]
    \centering
    \includegraphics[width=0.79\linewidth]{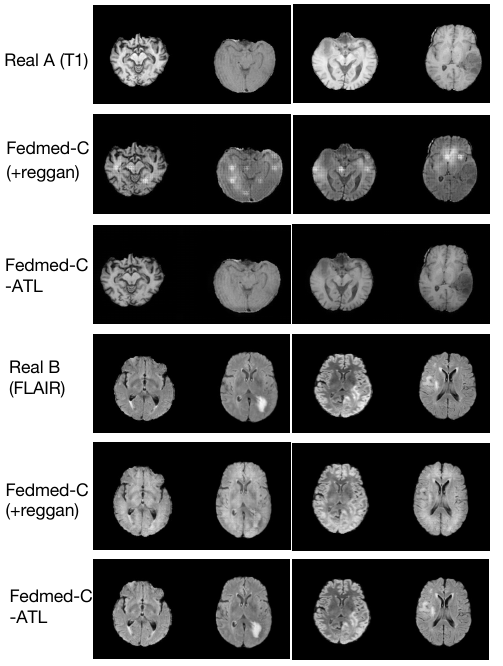}
	\caption{Visualization of brain images generated by FedMed-C~(+reggan) and FedMed-C-ATL on the BraTS validation set. The model is trained on the severe noise scheme. The row images represent two volume examples of two near slices. 
	}\label{fig:vis-brats}
\end{figure}

\begin{figure}[thbp]
    \centering
    \includegraphics[width=1\linewidth]{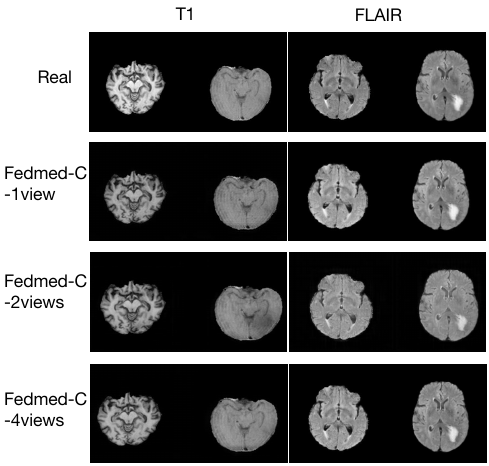}
	\caption{Visualization of brain images generated by FedMed-C-ATL-1view, FedMed-C-ATL-2views and FedMed-C-ATL-4Views on the BraTS validation set. The model is trained on the severe noise scheme. 
	}\label{fig:vis-num-view}
\end{figure}

\begin{figure}[thbp]
    \centering
    \includegraphics[width=1\linewidth]{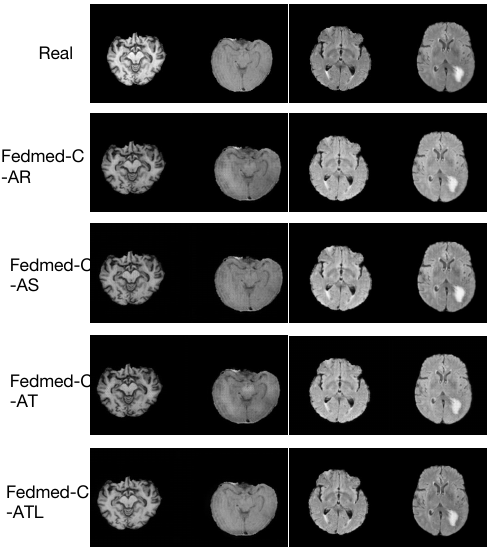}
	\caption{Visualization of brain images generated by FedMed-C-AR, FedMed-C-AT, FedMed-C-AS and FedMed-C-ATL on the BraTS validation set. The model is trained on the severe noise scheme. 
	}\label{fig:vis-ats}
\end{figure}

To present the performance of the synthesized images by our federated mode (FedMed), we provide the generated results of FedMed-C (+reggan) and FedMed-C-ATL in IXI (Figure~\ref{fig:vis-ixi}) and BraTS (Figure~\ref{fig:vis-brats}), the generated results of FedMed-C-ATL-1view, FedMed-C-ATL-2views, FedMed-C-ATL-4views in Figure~\ref{fig:vis-num-view} and the generated results of FedMed-C-AR, FedMed-C-AT, FedMed-C-AS and FedMed-C-ATL in Figure~\ref{fig:vis-ats}. From the results, we can see that FedMed-C-ATL has the ability to generate high-quality images, where the texture information of tissues and modal style would be well maintained. 

\section{Conclusions}
In practice, the misaligned and unpaired neuroimaging data (MUD) are inevitable, however, the traditional deformable registration methods require expensive computational resources. From our perspectives, MUD can be regarded as data augmentation, and can be treated as an important manner to improve the quality of the synthesized neuroimaging data. FedMed-ATL provides a simple but effective way to facilitate multi-modal brain image synthesis. We prove that FedMed-ATL outperforms state-of-the-art algorithm when data in severe MUD settings. Hopefully, FedMed-ATL can motivate the medical GAN community to focus on the severe MUD scenario.

\section{Acknowledgments}
This work is supported by the National Natural Science Foundation of China under Grant No. 61972188 and 62122035.

\bibliographystyle{ACM-Reference-Format}
\bibliography{sample-base}

\end{document}